# ROSAT PSPC observations of the infrared quasar IRAS 13349+2438: evidence for a warm absorber with internal dust

W.N. Brandt,[1] A.C. Fabian[1] and K.A. Pounds[2]

[1] *Institute of Astronomy, Madingley Road, Cambridge CB3 0HA (Internet: wnb@ast.cam.ac.uk, acf@ast.cam.ac.uk)*
[2] *X-ray Astronomy Group, Department of Physics & Astronomy, University of Leicester, University Road, Leicester LE1 7RH (Internet: kap@star.le.ac.uk)*

**ABSTRACT**

We present spatial, temporal and spectral analyses of *ROSAT* Position Sensitive Proportional Counter (PSPC) observations of the infrared loud quasar IRAS 13349+2438. IRAS 13349+2438 is the archetypal highly-polarized radio-quiet QSO and has an optical/infrared luminosity of $\approx 2 \times 10^{46}$ erg s$^{-1}$. We detect variability in the *ROSAT* count rate by a factor of 4.1 in about one year, and there is also evidence for $\approx 25$ per cent variability within one week. We find no evidence for large intrinsic cold absorption of soft X-rays. These two facts have important consequences for the scattering-plus-transmission model of this object which was developed to explain its high wavelength-dependent polarization and other properties. The soft X-ray variability makes electron scattering of most of the soft X-rays difficult without a very peculiar scattering mirror. The lack of significant intrinsic cold X-ray absorption together with the large observed $E(B-V)$ suggests either a very peculiar system geometry or, more probably, absorption by warm ionized gas with internal dust. There is evidence for an ionized oxygen edge in the X-ray spectrum. IRAS 13349+2438 has many properties that are similar to those of 'narrow-line' Seyfert 1s.

## INTRODUCTION

IRAS 13349+2438 is the first previously unidentified quasar that was selected through its infrared emission and is the prototype radio-quiet, infrared-bright quasar (Beichmann et al. 1986). It has a redshift of $z = 0.107$ and a high polarization percentage which rises towards the blue from 1.4 per cent at 2.2 $\mu$m ($K$-band) to 8 per cent at 0.36 $\mu$m ($U$-band) (Wills et al. 1992, hereafter W92). W92 found no variability of the polarization or flux density on time scales from days to months. To explain its polarization and other properties, W92 have developed an elegant model for the inner regions of the QSO and argue for a bipolar geometry. They suggest that a direct, but attenuated, AGN spectrum reaches Earth through a dusty inclined disc that is parallel to the plane of the host galaxy and probably between the broad line region and the narrow line region. In addition, light escaping in polar directions is much less attenuated and is scattered towards Earth. This scattered light is hence polarized parallel to the major axis of the host galaxy.

Beichmann et al. (1986), Barvainis (1987) and W92 have modelled the dust emission from IRAS 13349+2438. The optical continuum of IRAS 13349+2438 exhibits a distinct upturn in the blue above about 1 $\mu$m, in contrast to what is often seen in powerful infrared quasars (see sect. III.c of Frogel et al. 1989). Lanzetta, Turnshek & Sandoval (1993) present *IUE* spectra of IRAS 13349+2438, and comment that its ultraviolet continuum slope may change with time (see their sect. 3.1). Despite its high polarization and radio-quiet nature, IRAS 13349+2438 has not been seen to have broad absorption lines (BAL; see Boroson & Meyers 1992).

In this paper we present a detailed study of the soft X-rays from IRAS 13349+2438. We use the observed soft X-ray properties to explore and test facets of the W92 model. In particular, we shall examine soft X-ray absorption and shall compare our measured cold absorption with that suggested by the large reddening of $E(B-V) > 0.3$. *ROSAT* all-sky survey detections of IRAS 13349+2438 were briefly reported in Walter & Fink (1993, hereafter WF93) and Brinkmann & Siebert (1994), but our temporally separated and deeper pointed observations allow us to perform a much more sophisticated study of this important object. It is worth noting that IRAS 13349+2438 is one of the peculiar outlying objects in the WF93 *ROSAT* all-sky survey correlation between spectral slope and ultraviolet (1375 Å) to 2-keV flux.

A value of the Hubble constant of $H_0 = 50$ km s$^{-1}$ Mpc$^{-1}$ and a cosmological deceleration parameter of $q_0 = \frac{1}{2}$ have been assumed throughout.

## 2  OBSERVATIONS AND DATA REDUCTION

*ROSAT* PSPC (Trümper 1983; Pfeffermann et al. 1987) observations were made of IRAS 13349+2438 starting on 1992 Jan 7 (RP700553; total raw exposure of 3321 s over $5.99 \times 10^5$ s in two observation intervals) and 1992 Dec 16 (RP700553-1; total raw exposure of 1573 s in one observation interval). Hereafter we shall refer to RP700553 as 'Pointing 1' or 'P1' and RP700553-1 as 'Pointing 2' or 'P2'. IRAS 13349+2438 has 3230 and 6120 raw background subtracted counts in P1 and P2, respectively.

The *ROSAT* observations were performed in the standard 'wobble' mode; to avoid accidental shadowing of sources by the coarse wire grid which forms part of the PSPC entrance window support structure, *ROSAT* performs a slow dithering motion diagonal to the detector axes with a period of $\approx 400$ s and an amplitude of 3 arcmin.

The PSPC detector at the focus of the *ROSAT* X-ray telescope has a bandpass of $\sim 0.1$–$2.5$ keV over a $2°$-diameter field of view. Its energy resolution is $\Delta E/E = 0.43(E/0.93)^{-0.5}$ (FWHM), where $E$ is the photon energy measured in keV. The spatial resolution for the full bandpass is $\approx 25$ arcsec (FWHM) on-axis (Hasinger et al. 1992). The PSPC background count rate is very low, and is dominated by the cosmic ray induced particle background, scattered solar X-rays and the diffuse X-ray background. The rejection efficiency for the particle background is excellent, and the residual count rate due to cosmic rays is only $\sim 5 \times 10^{-6}$ count s$^{-1}$ arcmin$^{-2}$ keV$^{-1}$ (Snowden et al. 1992).

Reduction and analysis of the PSPC image data was performed with the Starlink ASTERIX X-ray data processing system.

## 3  ANALYSES

### 3.1  Spatial analysis

The 0.1–2.5 keV spatial profile of IRAS 13349+2438 in both observations is consistent with that of a point source convolved with the *ROSAT* XRT and PSPC spatial responses (Hasinger et al. 1992). IRAS 13349+2438 lies in the centres of the *ROSAT* fields of view for both observations. The X-ray centroid of the source we identify with IRAS 13349+2438 is $\alpha_{2000} = $ 13h37m18.8s, $\delta_{2000} = $ +24d23m11s during P1, and its centroid is $\alpha_{2000} = $ 13h37m18.9s, $\delta_{2000} = $ +24d23m07s during P2. These positions are consistent with the optical position of $\alpha_{2000} = $ 13h37m18.8s, $\delta_{2000} = $ +24d23m04s and the positional error of the *ROSAT* PSPC of $\approx 20$ arcsec.

We have extracted source counts for IRAS 13349+2438 from circular source cells chosen to be large enough to ensure that all of the source counts are included, given the electronic 'ghost imaging' which widens the point spread function below $\sim 0.3$ keV (Hasinger et al. 1992). The radii of the circular source cells were not a function of channel energy. Background counts were subtracted from the source cells using large nearby circular source-free background cells. Corrections were included for PSPC dead time, vignetting and shadowing by the coarse mesh window support. We have conservatively disregarded data from the first 40 s and last 10 s of each observation interval so as to avoid potential detector voltage and satellite drift problems. We use these data in the temporal and spectral analyses below.

### 3.2  Temporal analysis

Count rates should be averaged over the 400 s *ROSAT* wobble period when used for source flux determination, as variability on timescales shorter than this can be influenced by thick and thin wire shadowing. The wobble path can be geometrically distorted by digitization uncertainties in the star sensor read-out, and this can lead to apparent variations in count rate over different wobble periods. However, such effects are generally small and Brinkmann et al. (1994) state that the accuracy of flux determinations in wobble mode is good to within $\approx 4$ per cent. We have used 400-s binning for all temporal analysis work.

We have searched the *ROSAT* housekeeping data for instrumental effects that might produce the variability discussed below. The aspect uncertainty (`ASP-ERR`) is normal for all observations. We have removed data where the master veto rate `EE-MV` $> 115$. We have checked for general correlations between the count rate of IRAS 13349+2438 and the master veto rate, and no correlations are apparent. We have done this for both P1 and P2 jointly as well as for just P1 and just P2 (to prevent the relatively large source count rate change between P1 and P2 from confusing our correlation checking). We have examined the count rates of a serendipitous source near the centre of the *ROSAT* field of view, and it shows no signs of correlated variability with IRAS 13349+2438 between P1 and P2 (this source was the brightest one in the field after IRAS 13349+2438). The source is too faint to allow us to sensitively look for correlated variability within just pointing P1. It is located at $\alpha_{2000} = $ 13h38m07.8s, $\delta_{2000} = $ +24d24m17s. It is not listed in NED and has a count rate of about 0.04 counts s$^{-1}$. We have repeated some of the temporal analysis of the data using the IDL *ROSAT* analysis software system and obtain the same general results.

Fig. 1 shows the IRAS 13349+2438 full *ROSAT* band light curves during the P1 and P2 *ROSAT* pointings. The weighted mean count rate during P2 is a factor of 4.1 times that that during P1, and during P2 IRAS 13349+2438 has become one of the brighter AGN in the *ROSAT* sky. The end of the P1 pointing and the start of the P2 pointing are separated by about 337 days.

A comparison of the data from the first and second observation intervals of P1 suggests further variability. Exactly the same source and background regions were used for both observation intervals in making this comparison and both observation intervals were processed together. A constant model is a poor description of the P1 data giving $\chi^2_\nu = 3.3$ for seven degrees of freedom (we have included 4 per cent flux determination errors as per Brinkmann et al. 1994 in this analysis). This $\chi^2_\nu$ can be rejected at greater than 99 per cent confidence.

We have examined ratios of 0.5–2.5 keV count rate to 0.1–0.5 keV count rate and there is no strong evidence for hardness ratio variability either between P1 and P2 or between the two observation intervals of P1.

### 3.3  Spectral analysis

Counts from the corrected circular source cells were extracted into 256-channel, pulse-invariant spectra. We ignored channels 1–8 and rebinned the extracted spectra so that at least 100 source photons were present in each bin.

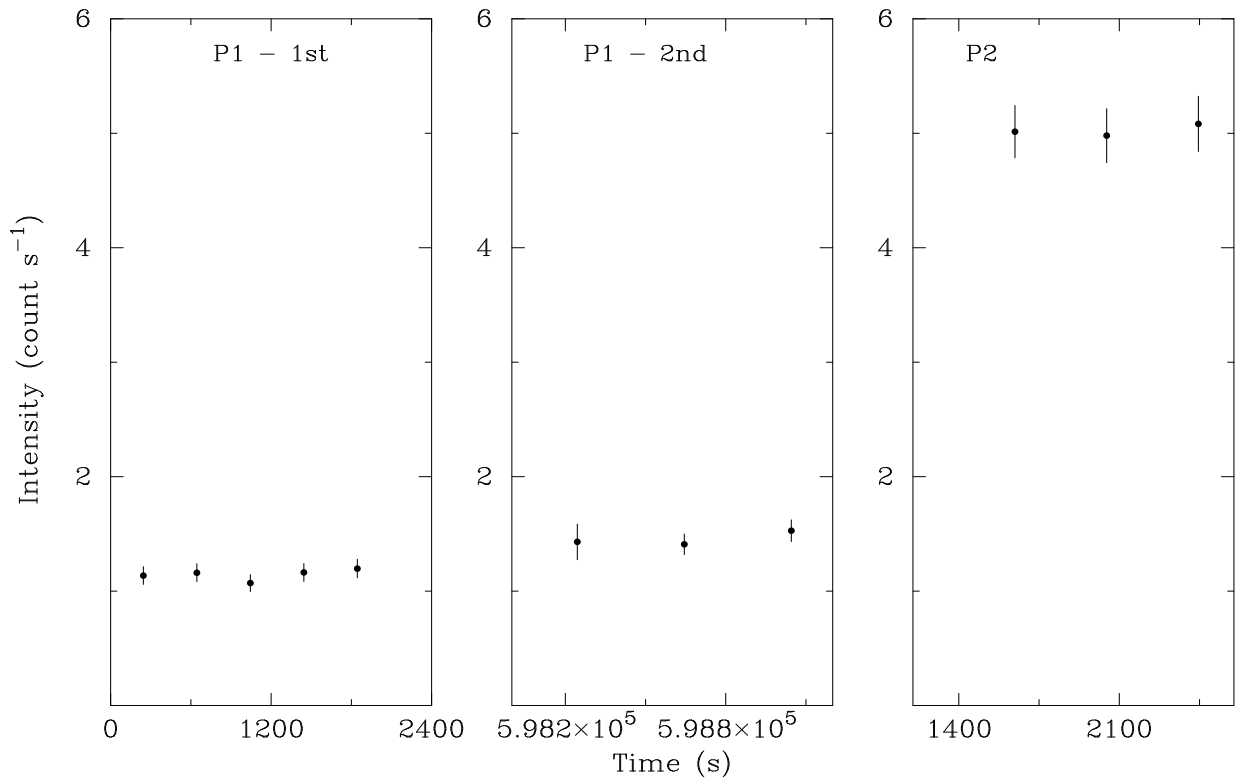

**Figure 1.** The *ROSAT* full band light curves of IRAS 13349+2438 for the first observation interval of P1 (left), the second observation interval of P1 (middle) and the P2 observation (right). The P1 and P2 observations have different time axis zero points, and times are measured in seconds from the starts of the P1 and P2 observations. The last data point in the left panel and the first data point in the middle panel are separated by about 6.9 days, and the last data point in the middle panel and the first data point in the right panel are separated by about 337 days. The data have been binned into 400-s bins (corresponding to the wobble period), and are corrected for detector dead time, vignetting and wire shadowing. The plotted error bars include the 4 per cent flux determination error mentioned in the text. IRAS 13349+2438 brightened between P1 and P2, becoming one of the brighter AGN in the *ROSAT* sky.

The widths of the bins that resulted were smaller than the PSPC's spectral resolution, and this oversampling obviates potential biasing problems that could arise as a result of sharp spectral changes. Rebinning the extracted spectra with the requirement that at least 40 source photons were present in each bin did not materially change our results below.

Systematic errors of 2 per cent were added in quadrature to the data point rms errors, to account for residual uncertainties in the spectral calibration of the PSPC. We have used the 1992 March response matrix (MPE No. 6) for the P1 spectral fitting and the 1993 January response matrix (MPE No. 36) for the P2 spectral fitting (these are the recommended matrices according to the PSPC team). These matrices correct for the systematic deficit of photons near the carbon edge of the PSPC detector that was present in earlier matrices (see Turner, George & Mushotzky 1993). The expected systematic errors from these matrices are a few per cent, and are significantly smaller than the relevant residuals in the data discussed below. We are keenly aware, however, of remaining spectral calibration uncertainties with the PSPC, and refer the reader to the discussions of this issue in appendix A of Brinkmann et al. (1994), and appendix A and appendix B of Fiore et al. (1994).

We model the X-ray spectra of IRAS 13349+2438 using the X-ray spectral models in the XSPEC spectral fitting package (Shafer et al. 1991). The errors for all fits shall be quoted for 90 per cent confidence (unless explicitly stated otherwise), conservatively taking all free parameters to be of interest other than absolute normalization (Lampton, Margon & Bowyer 1976; Press et al. 1989). According to Stark et al. (1992), the Galactic neutral hydrogen column density towards IRAS 13349+2438 is $N_H = 1.0 \times 10^{20}$ cm$^{-2}$. Sect. 3.1 of Laor et al. (1994) indicates that an appropriate Galactic $N_H$ uncertainty is $\Delta N_H = 2 \times 10^{19}$ cm$^{-2}$. A recent precise radio measurement of the Galactic column towards IRAS 13349+2438 gives $N_H = (1.07 \pm 0.1) \times 10^{20}$ cm$^{-2}$ (A. Laor and J. Lockman, pers. comm.). We shall first fit the data from the second *ROSAT* pointing (P2) since it has more counts and shall then fit the data from the first

**Table 1.** Basic spectral fitting to IRAS 13349+2438 during Pointing 2 (P2; RP700553-1).

| Model Name | $N_{\rm H}$ /($10^{20}$ cm$^{-2}$) | $A_1$ /($10^{-3}$ ph keV$^{-1}$ cm$^{-2}$ s$^{-1}$) | $\Gamma_1$ | Other Parameters | $\chi^2$/d.o.f.=$\chi^2_\nu$ |
|---|---|---|---|---|---|
| Power Law | $1.20^{+0.25}_{-0.22}$ | $4.05^{+0.28}_{-0.27}$ | $2.81^{+0.14}_{-0.13}$ | — | 56.30/36=1.56 |
| Power Law & Blackbody | $0.48^{+0.45}_{-0.40}$ | $3.21^{+0.74}_{-1.01}$ | $1.81^{+0.71}_{-0.97}$ | $K=(1.99^{+0.69}_{-1.07})\times 10^{-4}$ $kT=0.09^{+0.01}_{-0.01}$ keV | 40.06/34=1.18 |
| Power Law & Edge | $1.11^{+0.34}_{-0.29}$ | $5.36^{+0.96}_{-0.88}$ | $2.58^{+0.22}_{-0.20}$ | $\tau=0.99^{+0.71}_{-0.62}$ $E_{\rm Edge}=0.86^{+0.10}_{-0.10}$ keV | 34.98/34=1.03 |

All spectral models above are described in Shafer et al. (1991), and the spectral parameters are defined as follows:
$N_{\rm H}$ = equivalent hydrogen column in atom cm$^{-2}$; $A_1$ = normalization of power law in photon keV$^{-1}$ cm$^{-2}$ s$^{-1}$ at 1 keV; $\Gamma_1$ = photon index of power law; $K$ = dimensionless normalization of blackbody (we have used the 'bbody' XSPEC model; see Shafer et al. 1991 for a definition of the normalization's meaning); $kT$ = blackbody temperature in keV; $\tau$ = edge absorption depth at threshold; $E_{\rm Edge}$ = edge threshold energy (corrected for redshift).
In the last column 'd.o.f.' stands for 'degrees of freedom.'
All error bars are for 90 per cent confidence, assuming that all free parameters are of interest other than the absolute normalization.

pointing (P1) in light of what we have learned.

Spectral fits to the 0.1–2.5 keV spectrum of IRAS 13349+2438 during P2 are shown in Table 1 and Fig. 2. A simple power-law fit is poor, and systematic residuals are apparent above $\approx 0.6$ keV. It can be ruled out with greater than 98 per cent confidence. A power-law and blackbody fit is statistically acceptable but reduces the fit $N_{\rm H}$ value to where it is inconsistent with the Galactic column (we shall discuss this further below). Any cold absorption intrinsic to IRAS 13349+2438 would only exacerbate this discrepancy. In addition, weak systematic residuals remain visible above $\approx 0.6$ keV. A power-law and edge model is statistically acceptable, removes all systematic residuals and yields an edge energy (corrected for redshift) of $0.86^{+0.10}_{-0.10}$ keV. In the PSPC band, the dominant ionized absorber is oxygen, both because of its large abundance (Morrison & McCammon 1983) and its large photoionization cross-section (Daltabuit & Cox 1972). It has K edge energies ranging from 0.533 keV (for O I) to 0.871 keV (for O VIII), and the edge energy we fit is consistent with absorption by a combination of O VII (which has an edge energy of 0.739 keV) and O VIII (see equation 8 of Mewe & Schrijver 1978 for oxygen edge energies). Absorption by 'warm' ionized oxygen has been seen in several other AGN, and we will consider more sophisticated ionized gas absorption models below. Fitting a power-law with both a blackbody soft excess and an edge is an overparametrization of the *ROSAT* data (although it is probably physically realistic), but when it is done the presence of the edge is robust in the sense that the edge is strong while the blackbody normalization is greatly reduced. The parameters of the fit are poorly constrained, but for the record their best-fit values are $\Gamma = 2.79$, $N_{\rm H} = 1.2 \times 10^{20}$ cm$^{-2}$, $K = 5.8 \times 10^{-5}$, $kT = 0.27$ keV, $\tau = 2.81$ and $E_{\rm Edge} = 0.77$ keV with $\chi^2_\nu = 1.06$ for 32 degrees of freedom. We have checked whether the addition of an edge to the power-law and blackbody model of Table 1 is a statistically significant improvement using the $F$-test for the two additional parameters (see sect. 10-2 of Bevington 1969). We calculate $\Delta\chi^2 = 6.0$ and $F = 2.8$, which indicates that the addition of an edge is significant at $> 90$ per cent confidence but $< 95$ per cent confidence. For the sake of completeness we have also considered models with a power law and bremsstrahlung soft excess as well as a power law and narrow (50 eV) Gaussian emission line. A power-law and bremsstrahlung soft excess model is unstable and is statistically a poor fit. A power-law and Gaussian line is also statistically a poor fit, with $\chi^2_\nu = 1.5$. Furthermore, we have fixed the cold column at $1.7 \times 10^{21}$ cm$^{-2}$ (see Sect. 4.1 for our motivation for choosing this value) as well as $8.5 \times 10^{20}$ cm$^{-2}$ (half of $1.7 \times 10^{21}$ cm$^{-2}$) and searched without success for statistically acceptable models using, for example, double blackbody and blackbody+bremsstrahlung soft excess models. For example, a power-law plus double blackbody model (that also allows for the presence of an ionized oxygen edge) with $N_{\rm H}$ fixed at $8.5 \times 10^{20}$ cm$^{-2}$ yields $\chi^2 = 67.30$ for 30 degrees of freedom (compare this $\chi^2$ value with the values in Table 1). This $\chi^2$ value can be ruled out with greater than 99.9 per cent confidence, and systematic residuals are clearly visible at low energies. Deleting sets of data points at low energies shows that the small intrinsic cold column we measure is insensitive to individual data points but rather arises from the robust overall shape of the low energy X-ray spectrum.

Spectral fitting of a power law with cold absorption to IRAS 13349+2438 during P1 gives $\Gamma = 2.81^{+0.21}_{-0.19}$, $N_{\rm H} = (1.13^{+0.38}_{-0.33}) \times 10^{20}$ cm$^{-2}$ and $\chi^2_\nu = 1.30$ with 22 degrees of freedom. Note that these values are consistent with those obtained from fitting a power law with cold absorption to the P2 data. This value of $\chi^2_\nu$ can be rejected at the 84 per cent confidence level and edge-like residuals are still marginally visible. If an edge is included we obtain $\Gamma = 2.70^{+0.33}_{-0.31}$, $N_{\rm H} = (1.17^{+0.63}_{-0.47}) \times 10^{20}$ cm$^{-2}$, $\tau = 0.70^{+0.75}_{-0.64}$, $E_{\rm Edge} = 0.75^{+0.23}_{-0.29}$ (corrected for redshift) and $\chi^2_\nu = 1.15$ with 20 degrees of freedom (here we have quoted the edge parameters assuming two parameters to be of interest). Note again that the best fit edge parameters agree well with absorption by ionized oxygen.

For the reasons stated above we prefer the edge model to the blackbody soft excess model. However, it must be noted that while the model with a soft excess yields a fit $N_{\rm H}$ below the Galactic column, PSPC calibration uncertainties might account for this small ($\lesssim 5 \times 10^{19}$ cm$^{-2}$) discrepancy (it is crucial to note here that PSPC calibration uncertainties *cannot* account for the immensely larger observed versus expected $N_{\rm H}$ discrepancy we discuss in Sect. 4.1). Furthermore, X-ray absorption by cold gas and dust can be extremely complex (e.g. Czerny et al. 1995) and hence limitations in the cold absorption model we fit must be considered. Therefore we shall further examine both models in our discussion, and we note that our main results below are independent of the details of the spectral modelling.

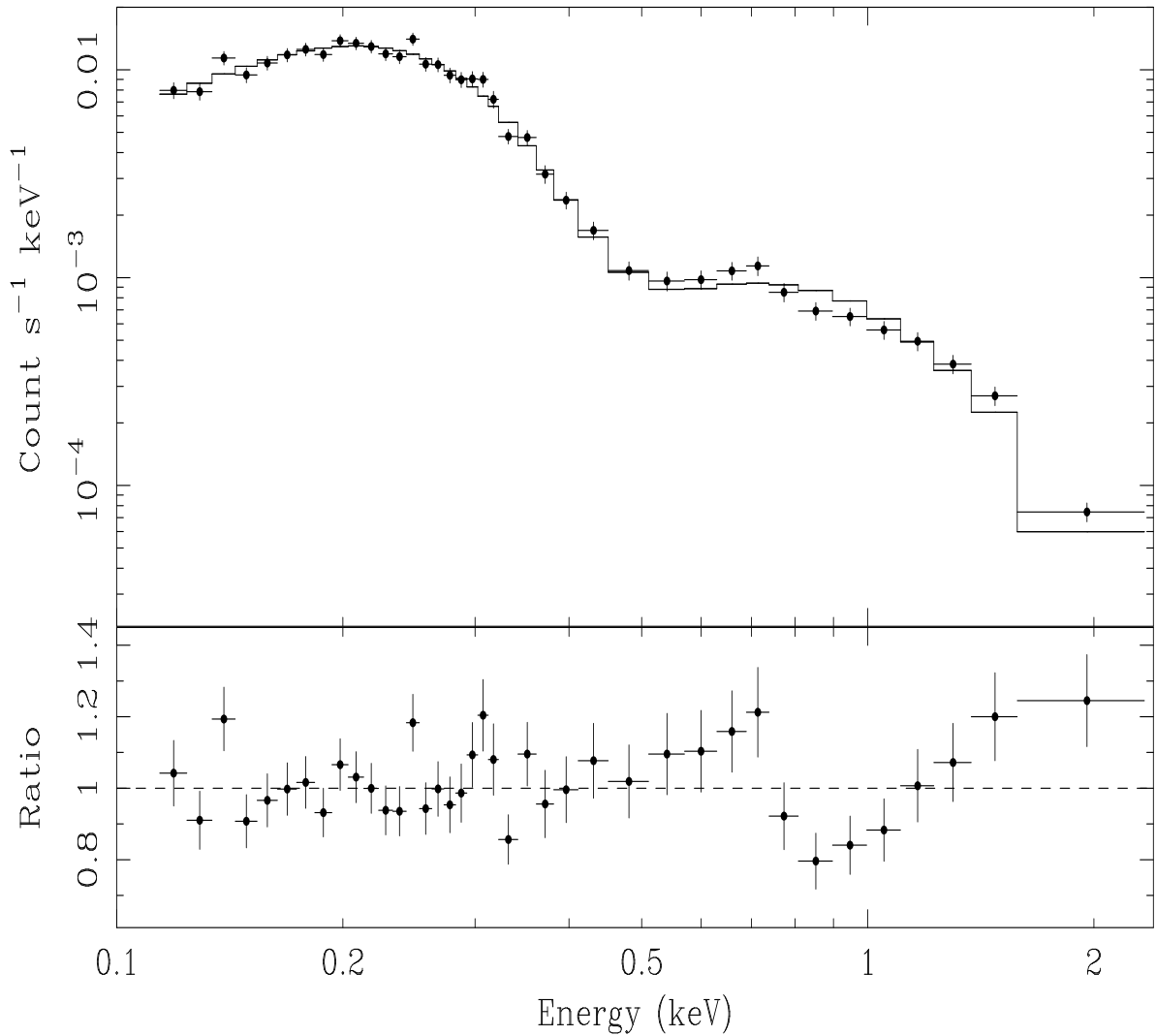

**Figure 2.** The *ROSAT* spectrum of IRAS 13349+2438 during the P2 pointing. The spectrum has not been corrected for redshift. A single power-law model fit and the data-to-model ratio are also shown. Note the systematic residuals above $\approx 0.6$ keV and the lack of a cold absorption drop-off at low energies.

Using the power-law model described above, the mean absorbed 0.1–2.5 keV X-ray flux for IRAS 13349+2438 during P1 is $5.5 \times 10^{-12}$ erg cm$^{-2}$ s$^{-1}$. This corresponds to a 0.1–2.5 keV unabsorbed flux of $1.2 \times 10^{-11}$ erg cm$^{-2}$ s$^{-1}$ and an isotropic luminosity of $5.9 \times 10^{44}$ erg s$^{-1}$ ($H_0 = 50$ km s$^{-1}$ Mpc$^{-1}$ and $q_0 = \frac{1}{2}$ as per Sect. 1). Using the power-law and edge model described above, the mean absorbed 0.1–2.5 keV X-ray flux for IRAS 13349+2438 is $5.7 \times 10^{-12}$ erg cm$^{-2}$ s$^{-1}$. This corresponds to a 0.1–2.5 keV unabsorbed flux of $1.2 \times 10^{-11}$ erg cm$^{-2}$ s$^{-1}$ and an isotropic luminosity of $5.9 \times 10^{44}$ erg s$^{-1}$.

Using the power-law and edge model described above, the mean absorbed 0.1–2.5 keV X-ray flux for IRAS 13349+2438 during P2 is $2.4 \times 10^{-11}$ erg cm$^{-2}$ s$^{-1}$. This corresponds to a 0.1–2.5 keV unabsorbed flux of $4.6 \times 10^{-11}$ erg cm$^{-2}$ s$^{-1}$ and an isotropic luminosity of $2.3 \times 10^{45}$ erg s$^{-1}$. Using the power-law and blackbody model described above, the mean absorbed 0.1–2.5 keV X-ray flux for IRAS 13349+2438 is $2.5 \times 10^{-11}$ erg cm$^{-2}$ s$^{-1}$. This corresponds to a 0.1–2.5 keV unabsorbed flux of $3.1 \times 10^{-11}$ erg cm$^{-2}$ s$^{-1}$ and an isotropic luminosity of $1.5 \times 10^{45}$ erg s$^{-1}$.

The *ROSAT* all-sky survey isotropic luminosity was $3.5 \times 10^{45}$ erg s$^{-1}$ (Brinkmann & Siebert 1994). The survey observation was taken over 1990 Dec 27–28, and a count rate has not been published for the survey observation. Note that the all-sky survey observation was made before either of our pointed observations and that IRAS 13349+2438 was brighter during this observation than during either of our pointed observations hence showing further X-ray variability.

In Fig. 3 we show the observed spectral energy distribution of IRAS 13349+2438 from radio to X-ray energies. The 0.34–100 $\mu$m optical/infrared isotropic luminosity of IRAS 13349+2438 is $\approx 2.3 \times 10^{46}$ erg s$^{-1}$ (Beichman et al. 1986 converted to $H_0 = 50$ km s$^{-1}$ Mpc$^{-1}$).

### 3.4 The Wide Field Camera (WFC) data

The *ROSAT* WFC all-sky survey data show a signal of 5.1 sigma in the WFC S1 (60–140 Å = 90–206 eV) band at $\alpha_{2000}$ = 13h37m21.3s, $\delta_{2000}$ = +24d22m58s. The corresponding 90 per cent WFC error circle has a radius of 43 arcsec. This source just failed to meet the criteria for inclusion in WFC 2RE Source Catalogue (Pye et al. 1995), where the intention was to have very few false sources. It is, however, included in Mason et al. (1995) where further WFC survey information on IRAS 13349+2438 may be found. During the pointed observations the WFC was less sensitive and IRAS 13349+2438 was not detected.

We note that the detection by the WFC independently argues for a small cold X-ray column.

## 4 DISCUSSION

### 4.1 X-ray absorption by cold gas

As stated in Sect. 3.3, the Galactic neutral hydrogen column density towards IRAS 13349+2438 is $N_H = (1.07 \pm 0.1) \times 10^{20}$ cm$^{-2}$. Stellar reddening studies (see sect. 4.1 of W92) indicate a Galactic $E(B-V) < 0.01$ towards IRAS 13349+2438. Our acceptable fits to the P1 *ROSAT* data all yield total $N_H$ values of less than $1.45 \times 10^{20}$ cm$^{-2}$ at 90 per cent confidence. These fits are sensitive to neutral gas both in atomic and molecular form. Thus we find an intrinsic cold column of less than $5 \times 10^{19}$ cm$^{-2}$ in IRAS 13349+2438 (redshift effects on the cold absorption do not change this result as we have verified using the 'zwabs' redshifted absorption model in XSPEC). The most probable intrinsic column suggested by our favoured power-law and edge model for the P2 observation is less than $1 \times 10^{19}$ cm$^{-2}$.

W92 have studied the IR-optical reddening intrinsic to IRAS 13349+2438 (see their sect. 4.3) and find an *apparent* reddening of $E(B-V) \approx 0.3$. When they interpret this in terms of their scattered spectrum plus transmitted spectrum model they derive an $E(B-V)$ for the transmitted spectrum in the range 0.3 to 0.7. If we assume a 'Galactic' dust-to-cold-gas ratio, the cold hydrogen column that corresponds to $E(B-V) = 0.3$ is $1.7 \times 10^{21}$ cm$^{-2}$ and the cold hydrogen column that corresponds to $E(B-V) = 0.7$ is $4.0 \times 10^{21}$ cm$^{-2}$ (see sect. VI of Burstein & Heiles 1978). The dust-to-cold-gas ratio appropriate for the centre of IRAS 13349+2438 is not known, and we shall examine this matter in more detail later. For comparison, Allen & Fabian (1992) recently found evidence for a roughly Galactic dust-to-cold-gas ratio in the obscured quasar 3C109. A similar situation exists for the low-redshift radio galaxy Cygnus A which has an intrinsic X-ray column of $(3.8^{+0.75}_{-0.71}) \times 10^{23}$ cm$^{-2}$ (Ueno et al. 1994) and $A_V \sim 50$ (Ward et al. 1991). Some of the metals associated with the hydrogen in the absorption column may be located in dust grains (see sect. 3.1.2 of Martin 1978). Sect. 7.3b of Spitzer (1978) suggests that dust grains contain about one-third of the C, N and O atoms in addition to most of the heavier element atoms in the Galactic interstellar medium (see also Sofia, Cardelli & Savage 1994). However, as long as these grains are not so large as to be individually optically thick to X-rays the amount of absorption by the grains is not much different than that by an equivalent mass of material in gaseous form (e.g. sect. 3.1.3 of Martin 1978). Thus, at first look, the intrinsic column suggested by the $E(B-V)$ data and the intrinsic column we measure appear to be *highly* inconsistent. Four possibilities that might resolve this large discrepancy suggest themselves:

1. The X-ray and IR-optical radiation do not travel through the same matter.
2. The gas and dust we are looking through in IRAS 13349+2438 is very different from the typical cold gas and dust in the Galactic interstellar medium. For example, the absorbing gas we are looking through could be highly ionized and/or the dust grains in the absorbing matter could be individually optically thick to X-rays and thus have large self-blanketing factors (see sect. III of Fireman 1974). The dust-to-cold-gas ratio in the centre of IRAS 13349+2438 could be very different from the Galactic one.
3. The obscuration is a strong function of time.
4. The methods used to derive $E(B-V) \approx 0.3$ are unreliable.

We shall examine the first two possibilities below, but address the other two here.

While we cannot strictly rule out changes in the obscuration between the observations of W92 and our observations, none of our spectral fits show evidence for changes in the column density. The WF93 column density is consistent with our values as well. However, we cannot rule out time dependent partial covering by very optically thick material.

W92 used the apparent reddening of the broad H$\beta$, H$\alpha$ and Pa$\alpha$ lines to derive $E(B-V) \approx 0.3$, given that the intrinsic line ratios for IRAS 13349+2438 are typical of those for luminous QSOs. This reddening correction, when applied to the continuum, made the corrected IR-optical spectrum match those of other QSOs. While broad line decrements must be treated with caution due to line transfer effects, collisional ionization and other uncertainties (see Gaskell & Ferland 1984; sect. 7.4 of Netzer 1990), the fact that the H$\alpha$/H$\beta$ and Pa$\alpha$/H$\beta$ ratios for IRAS 13349+2438 are among the largest observed for QSOs as well as its continuum shape strongly suggest that IRAS 13349+2438 does show a large amount of reddening. Additionally, its infrared, Fe II, Ca II and Na I emission suggest large amounts of cold gas and dust are present in the nucleus (see sect. 5.3 of W92; although these last indicators do not necessarily show that the cold gas and dust lie along the line of sight).

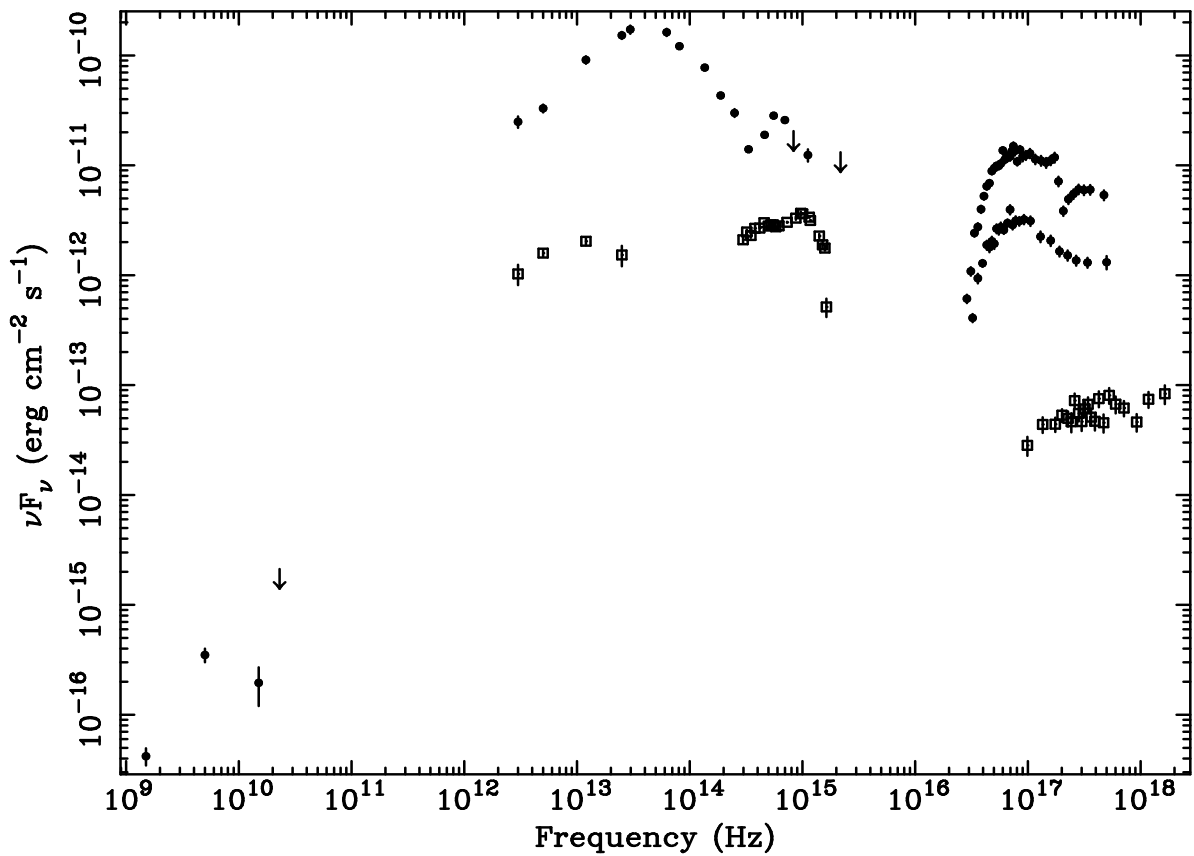

**Figure 3.** Spectral energy distributions of IRAS 13349+2438 (solid dots and arrows) and, for comparison, the radio quiet quasar PG 1634+706 (open squares). Spectral data for IRAS 13349+2438 are taken from Beichman et al. (1993), WF93 and our *ROSAT* pointings. We plot our *ROSAT* data from both the P1 and P2 pointings. The P1 data have been 'unfolded' through the PSPC response using the single power-law model fit and the P2 data have been 'unfolded' using the power-law and edge fit from Table 1. Spectral data for PG 1634+706 are taken from Nandra et al. (1995) and have been divided by a factor of 10 for clarity of presentation. Observations are not simultaneous. The strong flux decrease between $\sim 10^{15}$ Hz and $\sim 10^{17}$ Hz is due to photoelectric absorption by cold gas and reddening by dust. Note the giant infrared bump of IRAS 13349+2438.

### 4.2 The soft X-ray variability and X-ray scattering

The flux changes we see in IRAS 13349+2438 are somewhat surprising for such a powerful object. Studies of quasar soft X-ray variability (e.g. Zamorani et al. 1984; sect. 5.2 of Laor et al. 1994) generally find that the soft X-ray flux does not vary by more than a factor of two from the mean over timescales shorter than a few years. The X-ray variability power spectrum of IRAS 13349+2438 may flatten more gradually than for typical quasars. The fact that large flux changes are seen without strong spectral variability causes trouble for simple analogy schemes between Galactic black hole candidates and AGN where the soft flux would be expected to vary less (e.g. sect. 6.4 of Boller, Brandt & Fink 1995, hereafter BBF95; sect. 3 of Fiore & Elvis 1995). The variability and soft spectrum we see echoes the relation found in sect 3.4.2 of Green, McHardy & Lehto (1993).

X-ray scattering around the obscuring matter seen by W92 might allow X-rays to come to us without the signature of cold absorption. If the X-rays are scattered, this makes electrons the most probable scatterers (see sect. 5.2.2 of W92). However, the soft X-ray variability we see presents a challenge to models where *most* of the X-rays are scattered to us. The factor of 4.1 variability in about 337 days and causality arguments (assuming no beaming; note IRAS 13349+2438 is radio quiet) allow us to place an upper limit on the size of the region from which the X-rays come of $\approx 0.3$ pc. This upper limit could be increased somewhat if a small part of the region varied by a larger amount than the mean variation, but the large variability amplitudes implied soon become extreme for an object of this luminosity. Additionally, the ≈ 25 per cent variability suggests an X-ray source region with an extent of order 0.006 pc or less. These dimensions are to be compared with the size of the scattering mirror in the much less luminous Seyfert 2 galaxy NGC 1068, where photoionization modelling (Miller, Goodrich & Matthews 1991) and Hubble Space Telescope multiaperture spectropolarimetry (Antonucci, Hurt & Miller 1994) indicate a mirror size of ≈ 50 pc or more.

Another argument against scattering of *most* of the soft X-rays is the very large intrinsic soft X-ray luminosity or extreme mirror properties that would be required for IRAS 13349+2438 if its X-rays were predominantly scattered. We shall consider the P2 observation when the *ROSAT* isotropic luminosity was $\geq 1.5 \times 10^{45}$ erg s$^{-1}$ (see Sect. 3.3). The observed $L_{\rm X}/L_{\rm Bol}$ was $\geq \frac{1}{16}$ (we note $L_{\rm Bol}$ is dominated by the infrared luminosity which generally lacks significant time variability; Cutri et al. 1985; Neugebauer et al. 1989). The reflected fraction of incident flux by an electron scattering mirror, $f_{\rm refl}$, is approximately $\tau_{\rm T}(\Delta\Omega/4\pi)$, where $\Delta\Omega$ is the solid angle subtended by the mirror about the nucleus and $\tau_{\rm T}$ is its Thomson depth (corrections for the fact that Thomson scattering is dipolar and geometrical corrections such as those from sect. 4.5.1 of Miller, Goodrich & Matthews 1991 do not change the essential results). Thus unless $f_{\rm refl} \approx 1$, the $L_{\rm X}/L_{\rm Bol}$ ratio becomes extremely large for a radio-quiet AGN (see table 5 of Padovani & Rafanelli 1988). A value of $f_{\rm refl} \approx 1$ would be ≈ 66 times higher than that observed in NGC 1068 (Miller, Goodrich & Matthews 1991). For $f_{\rm refl}$ to be ≈ 1, this requires $\tau_{\rm T}$ to be ≈ 1 or more (if the opening half angle of the torus is ≈ 52 degrees as per sect. 5.1 of W92 then $\Delta\Omega/4\pi \approx 0.38$). This would be inconsistent with the optically thin mirror assumption of sect. 5.1 of W92 which was predicated on the observed high degree of polarization. A Thomson depth of ≈ 1 within 0.006 pc gives a fairly high minimum electron density of $\approx 8 \times 10^7$ cm$^{-3}$ for the scattering region. This density may be compared with the $\approx 8 \times 10^6$ cm$^{-3}$ upper limit on a typical Seyfert Broad Line Region (BLR) intercloud medium imposed by the fact that it must be Thomson thin (see sect. 2 of Perry & Dyson 1985; we conservatively consider a BLR intercloud medium with $\tau_{\rm T} = 0.2$ and a radius of $4 \times 10^{16}$ cm).

We therefore note that while some of the soft X-rays from IRAS 13349+2438 might be scattered by a mirror such as that in NGC 1068, a very atypical mirror which is both very compact and Thomson thick is required if most of its X-rays are scattered to us. The IR-optical flux may still be scattered by a mirror as per W92 (see below).

### 4.3 The X-ray data and IRAS 13349+2438 models

We now discuss the two possibilities of Sect. 4.1 that we did not elaborate on there.

#### 4.3.1 *Possibility 1: The IR-optical photons suffer extinction by dust but the X-rays escape through a 'borehole'*

The assumption that the X-rays and IR-optical emission travel through the same matter may be incorrect. This certainly seems likely to at least some degree in the inhomogeneous environment thought to exist around AGN nuclei. However, as we show in the next paragraph, explaining our X-ray observations in the context of the W92 model by invoking different X-ray and IR-optical light paths proves difficult.

X-rays and perhaps UV flux from the compact central source might escape through a 'borehole' of low extinction in a patchy torus (see sect. 5.1 of W92) while the more generally extended IR-optical emission does not. If the X-ray, UV, optical and IR emissions originate from successively greater radii of an accretion disc, a peculiar relatively perpendicular orientation of the disc and torus axes would seem to be required in this model (especially if the disc is geometrically thick). Such perpendicular orientations have not been identified in other objects. Comparisons of ionization cones and radio structures suggest that discs and tori may be highly coplanar (see sect. 4 of Wilson 1994 but see Mundell et al. 1995 for a tentative counterexample). The warm absorbing gas discussed above might fill the 'borehole' and the large flux variability (see Sect. 4.2) could arise via time dependent partial covering by a very thick torus cloud. If our line of sight grazes the edge of the torus as is suggested in W92, the 'borehole' might not be a hole but rather a serration in its edge (although again the geometry is difficult to envisage).

If one chooses to abandon the geometry of W92 then different X-ray and IR-optical light paths can be accommodated more easily. However, the price of this abandonment is a heavy one since the W92 model explains many of the properties of IRAS 13349+2438 with grace.

One possible alternative geometry postulates that there is reddening intrinsic to the source of optical photons. A thick dusty accretion disc might self-redden the photons it emits at large radii and wavelength dependent polarization could arise by transmission through dust (although W92 have shown that this dust would have to have a grain size distribution different from that in the interstellar medium of our Galaxy). However, such a model has serious difficulty explaining the similar polarization percentage of the H$\alpha$ line since this presumably comes from BLR clouds and not the disc itself. Furthermore, in such a model the fact that the polarization is parallel to the major axis of the host Galaxy must be taken to be a coincidence.

While we cannot strictly rule out even more complicated alternative geometries (e.g. a shell-like torus around the BLR with a tiny X-ray borehole that happens to lie along our line of sight), such models suffer from dynamical difficulties (e.g. it is unclear how a shell-like torus might be created and maintained), are generally anti-Copernican (e.g. they require tiny X-ray boreholes to be pointed right at Earth) and have difficulty naturally explaining why the polarization is parallel to the major axis of the host Galaxy.

#### 4.3.2 *Possibility 2: The gas and dust we are looking through in IRAS 13349+2438 is very different from the typical cold gas and dust in the Galactic interstellar medium*

Having found the other three possibilities of Sect. 4.1 incommodious, we consider the possibility that the gas and dust we are looking through in IRAS 13349+2438 is very different from that in the cold interstellar medium of our Galaxy. We shall first consider absorption by cold matter with an increased amount of dust per unit gas. We shall then con-sider the effects of gas ionization and shall finally consider the effects of changes in the dust grain size distribution.

Cold matter with a large dust-to-gas ratio could have a smaller $N_H/E(B-V)$ ratio than that of the local interstellar medium of our Galaxy. In this paragraph we shall consider the possibility that cold matter cloaking the nucleus of IRAS 13349+2438 has more dust per unit of gas. Even in the most conservative case, the dust content in IRAS 13349+2438 must be enhanced by a factor of $\gtrsim 34$ relative to that of our local interstellar medium. Given the same fraction of metal depletion into dust (see Sect. 4.1), this would require the metal abundance to also be enhanced by a factor of $\gtrsim 34$. Our X-ray spectral data cannot rule out such an enhancement (H and He dominate the fit column, see the next paragraph), but such an extreme metal abundance enhancement seems contrived. Even allowing for a total depletion of metals into dust, the required metal abundance enhancement is an extreme factor of $\gtrsim 11$. Total metal depletion into dust would make the ionized oxygen edge we see hard to explain.

Dust grains located in ionized rather than cold gas could explain the lack of large intrinsic cold X-ray absorption at energies below 0.5 keV without an extreme enhancement of metals. If the gaseous H, He, C, N, and O (or just some of these elements) were highly ionized this would reduce the cold absorption opacity significantly, especially at the important low energy end of the *ROSAT* band (see Krolik & Kallman 1984). We have examined the effects of ionization of the gaseous H and He by using the XSPEC 'fakeit' command to simulate *ROSAT* 0.1–2.5 keV observations of sources with power-law spectra of various photon indices absorbed by columns of $1.7 \times 10^{21}$ cm$^{-2}$ (see Sect. 4.1 to understand why we choose this value). However, in these spectra we take all of the H and He to be fully stripped (we use the power-law model with properly redshift-corrected variable-abundance absorption 'zvarabs' and set the H and He normalizations equal to zero to do this). We also include the cold Galactic column. We consider observations of the same length and count rate as the P2 observation. When we fit the resulting simulated spectra with a power-law plus standard cold column model, we obtain values of $N_H \sim 1 \times 10^{20}$ cm$^{-2}$ and generally acceptable values of $\chi^2_\nu$.

A model where we take a redshifted intrinsic column of $1.7 \times 10^{21}$ cm$^{-2}$ to have H and He fully stripped fits our P2 data well, provided that we also allow for the presence of an ionized oxygen edge. We include the Galactic column of $N_H = 1.07 \times 10^{20}$ cm$^{-2}$ in our fitting and obtain $\chi^2_\nu = 1.1$. Fit parameters for the power law and edge are poorly constrained but are not physically implausible.

In Sect. 4.1 we considered absorption by silicate and graphite dust with a composition and grain size distribution similar to that in the Galactic diffuse interstellar medium (e.g. Mathis, Rumpl & Nordsieck 1977). Deviations from a Mathis, Rumpl & Nordsieck (1977) dust grain size distribution in the vicinity of an AGN have been considered in sect. 7 of Laor & Draine (1993), and they find that a higher relative abundance of large grains is possible. Larger grains will reduce the extinction in the IR and optical (see fig. 6 of Laor & Draine 1993) and will also lead to self-blanketing which reduces the amount of X-ray photoelectric absorption by a given mass of dust (Fireman 1974). One might speculate that dust grains that can give an $E(B-V) \approx 0.3$ together with an $N_H \sim 1 \times 10^{20}$ cm$^{-2}$ are possible, even when the dust is located in cold gas. However, it must be remembered that H and He remain predominantly gaseous and these elements play a major role in determining the *ROSAT* band fit $N_H$. We have again used XSPEC to simulate spectra as in the previous paragraph but now take the column of $N_H = 1.7 \times 10^{21}$ cm$^{-2}$ to contain only H and He. Neglecting the other elements is equivalent to taking them all to be in dust grains with very effective self-blanketing factors. Fitting these spectra with a power-law plus standard cold column model, we obtain values of $N_H \sim 1 \times 10^{21}$ cm$^{-2}$, much larger than observed. Hence it seems unlikely that, by itself, a peculiar dust grain size distribution can explain our data. Ionization of at least the associated H and He appears to be required.

We have seen that the absorption in IRAS 13349+2438 below 0.5 keV may be explained by dusty gas in which at least H and He are ionized. In addition, at X-ray energies above 0.5 keV, we have independent evidence for a highly ionized oxygen edge. While association of these two gases as one and the same would be the simplest assumption and seems physically possible (see below), it must be noted that we cannot prove this assumption to be true. There could, for example, be dust-free gas very close to the central engine that produces the tentative oxygen edge as well as less highly ionized (but still ionized enough so that H and He are stripped) dusty warm gas further out. Below we shall state clearly whether we are talking about the tentative 'oxygen edge gas', the 'dusty warm gas' or both as one and the same.

### 4.4 X-ray absorption by warm gas

#### 4.4.1 Modelling of the oxygen edge gas

In this section we shall fit more sophisticated models to the structure in our data above 0.5 keV (see Fig. 2).

A one-zone warm gas absorption model constructed using the photoionization code CLOUDY (Ferland 1992; Reynolds et al. 1995) is a good fit to the P2 data with $\chi^2_\nu = 1.1$ for 34 degrees of freedom. The derived fit parameters are $\Gamma = 2.23^{+0.51}_{-0.69}$, $N_{H,\text{cold}} = (0.9^{+0.4}_{-0.4}) \times 10^{20}$ cm$^{-2}$, $\log(N_{H,\text{warm}}) = 22.78^{+1.20}_{-1.18}$ and $\xi = 145^{+80}_{-56}$. Note that the lower limit on $N_{H,\text{warm}}$ is larger than the $1.7 \times 10^{21}$ cm$^{-2}$ of Sect. 4.1 which may suggest a dust-to-gas ratio that is smaller than the Galactic one. However, limitations of current warm absorbing gas models must be kept in mind when the results from them are interpreted. For example, they assume strict photoionization equilibrium, thermal balance, and a simple gas density profile. Krolik & Kriss (1995) argue theoretically that all of these assumptions may well be invalid, and the results of Reynolds et al. (1995) hint that some are invalid as well.

The fact that the *ROSAT* isotropic luminosity of IRAS 13349+2438 varies by a large amount between P1 and P2 suggests that we may be able to learn the mode by which the gas that imprints the oxygen edge is ionized. In particular, if this gas were primarily photoionized and in photoionization equilibrium, we might expect significantly deeper edges in the P1 observation since the ionization parameter of the oxygen edge gas would be $\approx 3.9$ times less (this would be true only if the intrinsic isotropic luminosity of the source varied; changes only in the fraction of sky covered by very thick material would invalidate the argumentation of this paragraph). On the other hand, if collisional ionization were dominant or the warm absorbing gas were far from photoionization equilibrium, then we would not necessarily expect to see a change in edge depth. If we take the best fitting warm gas absorption model for P2 and allow only its normalization to change, we can fit the P1 data with $\chi_\nu^2 = 1.1$ for 24 degrees of freedom, a good fit. This is suggestive of a warm absorber that is not in simple photoionization equilibrium. To test this possibility further, we have fit the P1 data using the P2 model, allowing all parameters except $\xi$ and the power-law normalization to vary within their 90 per cent confidence regions from the P2 fit. $\xi$ and the power-law normalization are allowed to vary between their P2 fit lower bounds divided by 3.9 and their P2 fit upper bounds divided by 3.9. Unfortunately, we are not able to rule out this fit with statistical significance and therefore we cannot formally prove that the oxygen edge gas is not in simple photoionization equilibrium. Thus our spectral fitting constraints allow the oxygen edge gas to be either photoionized or collisionally ionized (but see the next section for further possible constraints).

We note that the ionized X-ray absorption we see is in line with the free-free radio absorption described by Beichman et al. (1986) as a way of explaining the sharply peaked radio spectrum of IRAS 13349+2438. Searches for sharply peaked radio spectra in other sources with prominent X-ray absorption edges (e.g. MCG$-6-30-15$, NGC 3227, NGC 3516) by ionized gas would be worthwhile.

### 4.4.2 *Constraints on the dusty warm gas*

The fact that we appear to see dust associated with at least some of the warm absorbing gas in IRAS 13349+2438 allows us to make inferences about the otherwise poorly constrained distance from the central engine and temperature of this gas.

Silicate dust grains sublime at $\approx 1400$ K and graphite grains sublime at $\approx 1750$ K. Laor & Draine (1993) state that the radius at which even large graphite dust grains exposed to an AGN spectrum sublime is $R_{\text{sublime}} \approx 0.20 L_{46}^{\frac{1}{2}}$ pc where $L_{46} = L_{\text{Bol}}/(10^{46}$ erg s$^{-1})$. Using this equation we derive $R_{\text{sublime}} \approx 0.3$ pc for IRAS 13349+2438. Netzer & Laor (1993) state that a good guess for the average BLR radius, as suggested by reverberation mapping studies, is $R_{\text{BLR}} \approx 0.06 L_{46}^{\frac{1}{2}}$ pc. Using this equation we derive $R_{\text{BLR}} \approx 0.09$ pc for IRAS 13349+2438. Thus it appears that the dusty warm gas we are looking through is probably located outside the BLR, perhaps near its outer edge (e.g. Netzer & Laor 1993; sect. 1 of Goodrich 1995). This is in agreement with the deductions of Barvainis (1987) and W92.

Thermal sputtering destroys internal dust once gas electron temperatures reach $10^6$ K (Draine & Salpeter 1979; sect. 7 of Laor & Draine 1993), so the dusty warm gas must have a gas electron temperature lower than this if the dust is going to exist for an extended period of time. The gas temperature in warm absorbing gas where oxygen is highly ionized is thought to be $\sim 10^5$ K if it is primarily photoionized and $\sim 10^6$ K or more if collisional ionization plays a significant role (e.g. Krolik & Kallman 1984). Thus, if the dusty warm gas and the oxygen edge gas are one and the same and the dust is not in the process of being destroyed by sputtering, the gas is probably photoionized rather than collisionally ionized.

### 4.4.3 *The warm dusty gas as a torus outflow or a 'torus atmosphere'*

If our line of sight to the central source in IRAS 13349+2438 grazed the edge of a torus, as suggested by W92, then we could be looking through dusty gas that is in the process of being torn from the torus and turned into an ionized outflow (see sect. III.a of Krolik & Begelman 1986). A geometrically thin torus or accretion disc with an outflow that blocks photons at high latitudes might simulate a geometrically thick torus (e.g. Königl & Kartje 1994). It could thereby match ionization cone and other observations and also avoid the dynamical difficulties associated with geometrical thickness. Thomson thick chunks of matter (perhaps clouds from the torus) in this outflow could lead to the X-ray variability we see as they pass through the line of sight (see Wachter, Strauss & Filippenko 1988). The ionized gas we see would then perhaps be the thermal wind described by Weymann, Turnshek & Christiansen (1985) to accelerate BAL clouds. Evidence for a warm absorbing outflows have been seen in other high luminosity objects such as 3C351 (Mathur et al. 1994). Alternatively, we could be looking through less rapidly moving gas in a 'torus atmosphere' (see fig. 1 of Tsvetanov, Kriss & Ford 1994). *ASCA* observations may be able to distinguish between these two possibilities by checking for shifts of the oxygen edge energies. In addition, *ASCA* will be able to look for temporal variability of oxygen edges and will be able to probe for edges from elements besides oxygen.

### 4.5 IRAS 13349+2438 and narrow-line Seyfert 1s

IRAS 13349+2438 appears to mix properties typically associated with Seyfert 1s, such as fairly rapid soft X-ray variability and little cold X-ray absorption, with properties typically associated with Seyfert 2s, such as high optical obscuration and wavelength dependent polarization. It has [O III] 5007 Å/H$\beta$=0.17, and Shuder & Osterbrock (1981) have shown that Seyferts with [O III] 5007 Å/H$\beta < 3$ are usually Seyfert 1's. It is worth seeking out other similar objects for comparison. The class of objects called 'narrow-line' Seyfert 1s (e.g. Osterbrock & Pogge 1985; Goodrich 1989; Puchnarewicz et al. 1992; BBF95; and references therein; hereafter NLS1) has at least some members that are similar. Mrk 766, for example, has both rapid soft X-ray variability (Molendi, Maccacaro & Schaeidt 1993) as well as a high optical polarization fraction ($\approx 2$ per cent; Goodrich 1989). Its optical polarization position angle is approximately perpendicular to its radio axis (Ulvestad, Antonucci & Goodrich 1995) and this is the typical orientation seen for Seyfert 2s. It deserves mention that Mrk 766, like IRAS 13349+2438, is one of the peculiar outlying objects in fig. 8 of WF93. It shares the $E(B-V)$ versus X-ray column discrepancy we see in IRAS 13349+2438 as does Ark 564, another NLS1 and WF93 outlying object (Brandt et al. 1994).

A comparison of the properties of IRAS 13349+2438 with those tabulated for NLS1 in the recent systematic study of 31 NLS1 by BBF95 reveals a strikingly large number of similarities:

1. A simple power-law fit to the *ROSAT* spectrum of IRAS 13349+2438 yields a steep photon index of $\Gamma = 2.81^{+0.14}_{-0.13}$.
2. IRAS 13349+2438 has rather rapid soft X-ray variability when compared to other objects of its luminosity.
3. IRAS 13349+2438 has lines from its high density region that are relatively narrow ($\approx 2100$ km s$^{-1}$; W92).
4. IRAS 13349+2438 has strong optical Fe II emission (W92).
5. IRAS 13349+2438 shows a deficit of ultraviolet emission relative to its soft X-ray spectrum (WF93).
6. IRAS 13349+2438 has weak narrow line emission (see fig. 2 of W92).

IRAS 13349+2438 satisfies the Goodrich (1989) criteria for classification as a NLS1 and when plotted in fig. 8 of BBF95 it lies among the NLS1. The strong Fe II emission and weak [O III] emission agree with the relationship embodied in the fundamental eigenvector of Boroson & Green (1992).

In light of the similarities of IRAS 13349+2438 and NLS1, it is worth investigating whether IRAS 13349+2438 might be profitably thought of as a NLS1 and whether the absorption physics we have used to explain IRAS 13349+2438 might apply to all NLS1. Absorption by warm gas can lead to steep NLS1 *ROSAT* band spectra (e.g. BBF95; Czerny et al. 1995). Dusty warm gas could lead to optical BLR lines with relatively small FWHM (by extinction) without large amounts of cold X-ray absorption (see sect 6.5 and fig. 8 of BBF95). It could also effectively absorb ultraviolet flux. This possibility may be further tested with infrared observations of NLS1 BLR lines to see if they are systematically broader than their optical BLR lines (due to the smaller infrared extinction). Dusty warm absorption does not seem to provide a natural explanation for the rapid flux and spectral variability of many NLS1 (BBF95; Otani 1995; but see Sect. 4.4.3 for interesting speculation), and warm absorption edges are seen in some (e.g. Ark 564, Brandt et al. 1994; NGC 4051, Mihara et al. 1994) but not all NLS1 (BBF95).

The well studied object MCG$-6-30-15$ suffers from a similar $E(B-V)$ versus $N_H$ discrepancy, shows an X-ray absorption edge largely due to ionized oxygen (e.g. Nandra & Pounds 1992), has a weak ultraviolet flux given the rest of its spectral energy distribution (fig. 8 of WF93; Reynolds & Fabian 1995) and has a high wavelength dependent polarization with a position angle parallel to the major axis of the host galaxy (Thompson & Martin 1988). It has $E(B-V) \approx 0.5$ (table 6 of Busko & Steiner 1990) and an X-ray measured intrinsic $N_H < 2 \times 10^{20}$ cm$^{-2}$. These similarities suggest that much of our dusty warm absorption discussion may be germane to MCG$-6-30-15$ as well.

## 5 SUMMARY

We have reported on *ROSAT* PSPC observations of the prototype IRAS quasar IRAS 13349+2438. Our main results are the following:

1. We detect *ROSAT* band flux variability by about a factor 4 in about 1 year, and also find evidence for $\approx 25$ per cent variability within one week. No strong spectral variability is apparent. The largest 0.1–2.5 keV luminosity we observe is $\approx 2 \times 10^{45}$ erg s$^{-1}$.
2. We do not detect large amounts of X-ray absorption by neutral matter. This is in contrast to what might be expected due to this object's large $E(B-V)$. We do find evidence for absorption by ionized oxygen.
3. The X-ray variability and large X-ray luminosity argue against having *most* of the X-rays be electron scattered towards Earth.
4. A 'warm absorber' that has internal dust can explain the optical and X-ray absorption properties of this source. We may be looking through an ionized outflow (perhaps a BAL QSO wind; IRAS 13349+2438 has many properties that are similar to those of BAL QSOs) or a torus atmosphere into the nucleus of this source.
5. IRAS 13349+2438 mixes properties typically associated with Seyfert 1's with properties typically associated with Seyfert 2's. It has a strikingly large number of similarities to 'narrow-line' Seyfert 1's. We examine these similarities and discuss their implications. The absorption properties of the well studied object MCG$-6-30-15$ may also be explained with a dusty warm absorber.

## ACKNOWLEDGMENTS

We thank S. Allen, M. Begelman, Th. Boller, A. Laor, K. Nandra, C. Otani, L. Puchnarewicz, C. Reynolds, M. Ward and B. Wills for useful discussions; H. Ebeling for help with IDL; and C. Reynolds for the use of his warm absorber models. We thank an anonymous referee for an illuminating and voluminous report. We gratefully acknowledge help from members of the Institute of Astronomy X-ray group and financial support from the United States National Science Foundation and the British Overseas Research Studentship Programme (WNB) and the Royal Society (ACF). The *ROSAT* project is supported by the Bundesministerium für Forschung und Technologie (BMFT) and the Max-Planck society. Much of our analysis has relied on the Starlink ASTERIX X-ray data processing system and the XSPEC X-ray spectral fitting software, and we thank the people who have created and maintain this software. This research has made use of data obtained from the UK *ROSAT* Data Archive Centre at the Department of Physics and Astronomy, University of Leicester and has also made use of the NASA/IPAC extragalactic data base (Helou et al. 1991) which is operated by the Jet Propulsion Laboratory, Caltech.

This paper has been produced using the Blackwell Scientific Publications TEX macros.